\documentclass[conference]{IEEEtran}
\IEEEoverridecommandlockouts
\usepackage[utf8]{inputenc} 
\usepackage[T1]{fontenc}
\usepackage{url}              
\usepackage{cite}             
\usepackage{float}
\usepackage[cmex10]{amsmath}  
\usepackage{mleftright}       
\mleftright                   
\usepackage{graphicx}         
\usepackage{booktabs}         
\usepackage{amsmath,amssymb,amsfonts}
\usepackage{algorithmic}
\usepackage{graphicx}
\usepackage{textcomp}
\usepackage{float}
\usepackage[dvipsnames]{xcolor}
\usepackage{enumerate}
\usepackage{algorithm}
\usepackage{bm}
\allowdisplaybreaks

\newtheorem{lemma}{Lemma}
\newtheorem{example}{Example}
\newtheorem{theorem}{Theorem}

\newtheorem{solution-to-dp}{Solution}
\newtheorem{computation assignment}{Computation Assignment}
\newtheorem{definition}{Definition}

\def\BibTeX{{\rm B\kern-.05em{\sc i\kern-.025em b}\kern-.08em
    T\kern-.1667em\lower.7ex\hbox{E}\kern-.125emX}}

\makeatletter
\newcommand\fs@spaceruled{\def\@fs@cfont{\bfseries}\let\@fs@capt\floatc@ruled
  \def\@fs@pre{\vspace{0.5\baselineskip}\hrule height.8pt depth0pt \kern2pt}%
  \def\@fs@post{\kern2pt\hrule\relax}%
  \def\@fs@mid{\kern2pt\hrule\kern2pt}%
  \let\@fs@iftopcapt\iftrue}
\makeatother

\begin{document}

\title{Uncoded Storage Coded Transmission Elastic Computing with Straggler Tolerance in Heterogeneous Systems}

\author{
\IEEEauthorblockN{Xi Zhong\textsuperscript{1}, Jörg Kliewer\textsuperscript{2} and Mingyue Ji\textsuperscript{1}}

\IEEEauthorblockA{\textit{\textsuperscript{1}Department of Electrical and Computer Engineering}, \textit{University of Utah}, Salt Lake City, UT, USA\\
Email: \{xi.zhong, mingyue.ji\}@utah.edu }

\IEEEauthorblockA{\textit{\textsuperscript{2}Department of Electrical and Computer Engineering}, \textit{New Jersey Institute of Technology}, Newark, NJ, USA\\
Email: jkliewer@njit.edu }}

\maketitle

\begin{abstract}
In 2018, Yang \emph{et al.} introduced a novel and effective approach, using maximum distance separable (MDS) codes, to mitigate the impact of elasticity in cloud computing systems. 
This approach is referred to as coded elastic computing. 
Some limitations of this approach 
include that it assumes all virtual machines have the same computing speeds and storage capacities, and it cannot tolerate stragglers for matrix-matrix multiplications.
In order to resolve these limitations, in this paper, we introduce a new combinatorial optimization framework, named uncoded storage coded transmission elastic computing (USCTEC), for heterogeneous speeds and storage constraints, aiming to minimize the expected computation time for matrix-matrix multiplications, under the consideration of straggler tolerance.  
Within this framework, we propose optimal solutions with straggler tolerance under relaxed storage  constraints. Moreover, we propose a heuristic algorithm that considers the heterogeneous storage constraints.
Our results demonstrate that the proposed algorithm outperforms baseline solutions utilizing cyclic storage placements, in terms of both expected computation time and storage size.
\end{abstract}

\section{Introduction} 
Elasticity allows virtual machines in a cloud system to be preempted or become available during computing rounds, leading to computation failure or increased computation time. 
In \cite{yang2018coded}, the authors proposed a cyclic computation assignment that utilizes maximum distance separable (MDS) coded storage for homogeneous systems, where all machines have the same computation speed and storage capacity. 
For MDS coded storage elastic computing, the authors in \cite{wcj2021hs} introduced a combinatorial optimization approach aimed at minimizing overall computation time for systems with heterogeneous computing speeds and storage constraints. They proposed an optimal solution using a low-complexity iterative algorithm, called the filling algorithm. 
Subsequently, in \cite{myjpractice}, the author extended the filling algorithm to address scenarios with both elasticity and stragglers.
In \cite{KSA2021}, the authors introduced two hierarchical schemes designed to speed up computing and tolerate stragglers, by letting fewer machines select their first computation tasks to work on and more machines select their last computation tasks.
In \cite{DHGR2020}, a new metric named transition waste was introduced, quantifying unnecessary changes in computation tasks caused by elasticity. 
To mitigate this, the authors minimized the transition waste among all cyclic computation assignments and constructed several computation assignments that achieve zero transition waste.

Despite the advantages of MDS coded storage elastic computing, they are limited to certain types of computations, such as linear computations. 
To overcome this limitation, the authors in \cite{usutec2022} introduced uncoded storage uncoded transmission elastic computing for heterogeneous systems. 
They formulated a combinatorial optimization problem and derived optimal solutions with the goal of minimizing the overall computation time for a given storage placement.

Most of the existing works in elastic computing, including \cite{yang2018coded, wcj2021hs, myjpractice, DHGR2020,usutec2022}, 
primarily focus on matrix-vector multiplications and utilize uncoded transmission during the communication phase.  
In \cite{yangCEC}, the authors proposed a coded storage coded transmission elastic computing scheme for matrix-matrix multiplications. However, this scheme cannot tolerate stragglers, as the MDS coded storage placement and transmission fix the number of machines contributing to the decoding process.

In this paper, we introduce the \underline{u}ncoded \underline{s}torage \underline{c}oded \underline{t}ransmission \underline{e}lastic \underline{c}omputing (USCTEC)
for systems with heterogeneous computation speeds and storage constraints. 
We first formulate a new optimization framework aimed at minimizing the expected computation time over a random distribution of computation speeds, using Lagrange codes, introduced in \cite{yu2019lagrange}, to design coded transmission and computation. 
Next, we design optimal USCTEC schemes with straggler tolerance, given any computation speed and no storage constraints. In this design, each machine stores a fraction of dataset.
Furthermore, we propose a heuristic algorithm that considers storage constraints for general speed distributions.
Finally, our results show that the proposed algorithm outperforms baseline algorithms that utilize cyclic storage placement, in terms of both expected computation time and required storage size.

\paragraph*{Notation}
$\mathbb{F}$ denotes a finite field, and $\mathbb{R}$ denotes the real field. 
We use $|\cdot|$ to represent the cardinality of a set or the length of a vector, and $[n] = \{1,2,\ldots,n\}$.
Let $a[i]$ denote the $i$-th element of vector $\boldsymbol{a}$, $\mu[i,j]$ denote the entry $[i,j]$ of matrix $\boldsymbol{\mu}$, and $\boldsymbol{\mu}[i]$ denote the $i$-th row of  $\boldsymbol{\mu}$.
We use $(\boldsymbol{B})_{\mathcal{D}}$ to represent the sub-matrix of $\boldsymbol{B}$ with column indices $\mathcal{D}$. 

\section{System Model and Problem Formulation}
\label{sec-network}
We consider a distributed system consisting of a master node and $N$ virtual machines, denoted by $[N]$. 
The computation speed is represented by a random vector $\boldsymbol{\mathsf{s}} = (\mathbf{s}[1], \cdots \mathbf{s}[N])$, where $\mathbf{s}[n]$ represents the number of row-column multiplications that machine $n$ can compute per unit of time. 
The sample space of the speed distribution is denoted as $\Omega_{\textbf{s}}$. 
Given a data matrix $\boldsymbol{A} \in {\mathbb{F}}^{q \times v}$, at each time step $t$, with the computation speed realization $\boldsymbol{s}^{(t)} \in \Omega_{\textbf{s}}$ and the input matrix $\boldsymbol{B}^{(t)} \in {\mathbb{F}}^{v \times r}$, a set of $ N_t$ available machines, known in the beginning of each step time and denoted as $\mathcal{N}_t = \{ n \in [N] : s^{(t)}[n] > 0\}$, aims to recover $\boldsymbol{A} \boldsymbol{B}^{(t)}$ while tolerating up to $S$ stragglers.
Define $L$ as the recovery threshold, which is the minimum number of machines required for successful decoding.
In the following, we explain how a USCTEC system operates. 

\subsection{Storage Placement and Storage Selections}
\label{subsec-storage}
Each machine $n \in [N]$ stores a subset of rows of the data matrix $\boldsymbol{A}$, denoted by $\mathcal{Z}_n$. The storage placement of the system is denoted by $\boldsymbol{\mathcal{Z}} =$ $\{\mathcal{Z}_{n}:$ $ n\in[N]\}$.
The storage constraint is presented by a vector $\boldsymbol{e} =$ $ (e[1]$, $\cdots$, $e[N])$, where $0 \leq e[n]$ $\leq 1$ for $n \in [N]$, 
and $e[n]$ indicates the maximum storage size of machine $n$, normalized by the size of  $\boldsymbol{A}$, i.e., $\frac{|\mathcal{Z}_n|}{q} \leq e[n]$. 

In each time step $t$, machine $n \in \mathcal{N}_t$ selects a subset of its storage $\mathcal{I}^{(t)}_n \subseteq \mathcal{Z}_n$ for computation tasks. 
Let $\boldsymbol{\mathcal{I}}^{(t)} = \{\mathcal{I}^{(t)}_n : n\in \mathcal{N}_t\}$. 
We obtain a specific $\boldsymbol{\mathcal{I}}^{(t)}$ by generating a partitioning vector $\boldsymbol{\gamma}^{(t)}$ and a set $\boldsymbol{\mathcal{U}}^{(t)}$. Specifically, $\boldsymbol{\gamma}^{(t)} =$ $(\gamma^{(t)}[1]$, $\cdots$, $\gamma^{(t)}[G^{(t)}])$ partitions $\boldsymbol{A}$ into 
$G^{(t)}$ disjoint row blocks, denoted as $\boldsymbol{A} =$ $\{\boldsymbol{A}_{g} \in$  ${\mathbb{F}}^{q \gamma^{(t)}[g] \times v} :$ $g$ $\in$ $[G^{(t)}]\}$, where $\sum_{g \in[G^{(t)}]}\gamma^{(t)}[g] = 1 $ and $ 0 $ $<$ $\gamma^{(t)}[g]$ $\leq 1$ for $g \in$ $[G^{(t)}]$. 
Next, we generate $\boldsymbol{\mathcal{U}}^{(t)} $ $= \{ \mathcal{U}^{(t)}_g:$ $g \in$ $[G^{(t)}]\}$. Each $\mathcal{U}^{(t)}_{g}$ is denoted as the selected machines for $\boldsymbol{A}_{g}$, where $\mathcal{U}^{(t)}_{g} \subseteq \mathcal{N}_{t}$, $|\mathcal{U}^{(t)}_{g}|\geq L+S$ and each machine in $\mathcal{U}^{(t)}_{g}$ stores $\boldsymbol{A}_{g}$.
Hence, the storage selection for machine $n$ is obtained by
\begin{equation}
    \label{eq-storage-selection}
    \mathcal{I}^{(t)}_n =\{\boldsymbol{A}_g : n\in \mathcal{U}^{(t)}_g, g \in [G^{(t)}]\}.
\end{equation}

Note that $\boldsymbol{B}^{(t)}$, $\boldsymbol{\mathcal{I}}^{(t)}$, $\boldsymbol{\gamma}^{(t)}$ and $\boldsymbol{{\mathcal{U}}}^{(t)}$ may change with each time step, but for simplicity, we omit the reference to the time step $t$ and denote $(\cdot)^{(t)}$ as $(\cdot)$. 

\subsection{Communication Phase}
The master partitions matrix $\boldsymbol{B}$ into $L$ blocks of equal size, denoted as $\boldsymbol{B}=$ $\{\boldsymbol{B}_{l} \in  {\mathbb{F}}^{v \times \frac{r}{L}} : l \in [L]\}$. Each $\boldsymbol{B}_{l}$ consists of $\frac{r}{L}$ columns, indexed by $[\frac{r}{L}]$. 
As a result, $\boldsymbol{A}\boldsymbol{B}$ consists of $G$ sets of blocks $\{\boldsymbol{A}_{g}\boldsymbol{B}_{l}: l\in[L]\}$ for $g \in [G]$. Each set will be recovered by the computation results from selected machines $\mathcal{U}_{g}$. 
To assign computation tasks to 
$\mathcal{U}_{g}$ for all $g \in [G]$, we 
define the computation assignment $\boldsymbol{\mathcal{M}}$.
\begin{definition}
    \label{def-assignment} 
    {\bf(Computation Assignment)} The computation assignment of the system is $\boldsymbol{\mathcal{M}} = \{(\boldsymbol{\mathcal{M}}_g, \boldsymbol{\mathcal{P}}_g): g\in[G]\}$, where the pair $(\boldsymbol{\mathcal{M}}_g, \boldsymbol{\mathcal{P}}_g)$ is the computation assignment for machines in $\mathcal{U}_g$. 
    $\boldsymbol{\mathcal{M}}_g = \{\mathcal{M}_{g,f} : f \in [F_g]\}$ represents an $F_g$-partition of the column indices $[\frac{r}{L}]$, i.e., $\bigcup_{f \in[F_{g}]}\mathcal{M}_{g,f} = [\frac{r}{L}]$.
    $\boldsymbol{\mathcal{P}}_{g} = \{\mathcal{P}_{g,f}: f \in [F_g]\}$ consists of $F_g$ sets of machines, where $\mathcal{P}_{g,f} \subseteq\mathcal{U}_{g}$ and  $|\mathcal{P}_{g,f}| = L+S$. 
    We denote that the machines in $\mathcal{P}_{g,f}$ are assigned to the indices $\mathcal{M}_{g,f}$, as they will be assigned to computation tasks associated with the columns in $\boldsymbol{B}_{l}$ with indices $\mathcal{M}_{g,f}$, for all $l \in [L]$.
\end{definition}

Based on 
$\boldsymbol{\mathcal{M}}$, the indices assigned to machine $n \in [N]$ are denoted as $\mathcal{D}_{n,g} = \bigcup_{f\in[F_g]: n \in \mathcal{P}_{g,f}} \mathcal{M}_{g,f}$ if $n \in \mathcal{U}_g$; otherwise, $\mathcal{D}_{n,g} = \emptyset$.
The overall assigned indices for machine $n$ are $\mathcal{D}_n = \bigcup_{g\in[G]} \mathcal{D}_{n,g}$.
To generate coded matrices for transmission, we use Lagrange codes introduced in \cite{yu2019lagrange}.
due to the low complexity and the capacity of straggler tolerance.
Specifically, the master selects $L$ numbers $\{\beta_{l} \in\mathbb{F}: l \in [L]\}$ and $N_{t}$  numbers $\{\alpha_{n} \in \mathbb{F}: n \in \mathcal{N}_t\}$ such that $\{\alpha_{n}: n \in \mathcal{N}_t\} \cap \{\beta_{l}: l \in [L]\} = \emptyset$. The master computes and sends the following coded matrix to machine $n \in \mathcal{N}_t$,
\begin{equation}\label{encoder}
    \boldsymbol{\Tilde{B}}_{n} = \sum_{l \in [L]} (\boldsymbol{B}_{l})_{\mathcal{D}_n }
    \cdot \prod_{k \in [L] \setminus {\{l\}}} {\frac{\alpha_{n}-\beta_{k}}{\beta_{l}-\beta_{k}}}.
\end{equation}

\subsection{Computing Phase and Decoding Phase} 
For $g \in[G]$,  machine $n \in \mathcal{U}_{g}$ computes and sends the following matrix to the master,
\begin{equation}
\label{eq-computing}
    \boldsymbol{H}_{g,n} = \boldsymbol{A}_{g} (\Tilde{\boldsymbol{B}}_n)_{\mathcal{D}_{n,g}}.
\end{equation}

For each block $\boldsymbol{A}_{g}\boldsymbol{B}_{l}$, $l\in[L]$, the master
decodes sub-block $\boldsymbol{A}_{g}(\boldsymbol{B}_{l})_{\mathcal{M}_{g,f}}$, using the computation results from machines in $\mathcal{P}_{g,f} \subseteq\mathcal{U}_{g} $. To do this, we define $F_g$ polynomials $H_{g,f}(z)$ with a degree of $L-1$ for $f\in[F_g]$, where
\begin{equation}
    \label{eq-outterpoly}
    H_{g,f} (z) = \boldsymbol{A}_{g} \cdot V_{g,f}(z), 
\end{equation}
\begin{equation}
    \label{eq-innerpoly}
    V_{g,f}(z) = \sum_{l \in [L]} (\boldsymbol{B}_{l})_{\mathcal{M}_{g,f} }
    \cdot \prod_{k \in [L] \setminus {\{l\}}} {\frac{z-\beta_{k}}{\beta_{l}-\beta_{k}}}.
\end{equation}
For each $H_{g,f}(z)$, $f \in[F_g]$,  we have two observations.
First, from \eqref{eq-innerpoly}, we have $V_{g,f} (\beta_{l})  = (\boldsymbol{B}_{l})_{\mathcal{M}_{g,f} }$ for $l\in[L]$. From \eqref{eq-outterpoly}, we have $ H_{g,f} (\beta_{l}) = \boldsymbol{A}_{g} (\boldsymbol{B}_{l})_{\mathcal{M}_{g,f} } $, i.e., the sub-block
is the evaluation of the polynomial $H_{g,f}(z)$ at $\beta_{l}$.
Second, due to $\mathcal{M}_{g,f} \subseteq \mathcal{D}_{n,g} \subseteq \mathcal{D}_{n}$, from \eqref{encoder} and \eqref{eq-innerpoly}, we have
\begin{equation}
\label{eq-midd}
    (\boldsymbol{\Tilde{B}}_{n})_{\mathcal{M}_{g,f} } 
    = V_{g,f} (\alpha_{n})  
\end{equation} 
for all $n \in \mathcal{P}_{g,f}$.
Then, $H_{g,f} (\alpha_{n}) \overset{(a)}{=}$ $\boldsymbol{A}_{g}V_{g,f} (\alpha_{n}) \overset{(b)}{=} \boldsymbol{A}_{g}(\boldsymbol{\Tilde{B}}_{n})_{\mathcal{M}_{g,f}} \overset{(c)}{=}  (\boldsymbol{H}_{g,n})_{\mathcal{M}_{g,f}}$,
where $(a)$ is due to \eqref{eq-outterpoly}, $(b)$ is due to \eqref{eq-midd} and $(c)$ is due to \eqref{eq-computing}.
In other words, the sub-matrix of computation result, i.e., $(\boldsymbol{H}_{g,n})_{\mathcal{M}_{g,f}}$, is the evaluation of the polynomial $H_{g,f}(z)$ at $\alpha_n$.
Therefore, decoding $\boldsymbol{A}_{g} (\boldsymbol{B}_{l})_{\mathcal{M}_{g,f} }$
for $l\in[L]$ and $f\in[F_g]$ means interpolating the polynomial $H_{g,f}(z)$ using the computation results $(\boldsymbol{H}_{g,n})_{\mathcal{M}_{g,f}}$ from any $L$ machines in $\mathcal{P}_{g,f}$, denoted by $\mathcal{L}_{g,f}$, and evaluating $H_{g,f}(\beta_{l})$.  
Using Lagrange interpolation, the master computes
\begin{equation*}
    \label{eq-decode}
     H_{g,f}(\beta_{l}) \!=\!\!\!\!\!\sum_{n\in \mathcal{L}_{g,f}} \!\!(\boldsymbol{H}_{g,n})_{\mathcal{M}_{g,f}} \cdot \!\!\!\!\!\!\! \prod_{n'\in \mathcal{L}_{g,f} \setminus \{n\}} \!\!\!\! \frac{\beta_{l}-\alpha_{n'}}{\alpha_{n}-\alpha_{n'}} \!= \!\boldsymbol{A}_{g} (\boldsymbol{B}_{l})_{\mathcal{M}_{g,f} }.
\end{equation*}

By combining $\boldsymbol{A}_{g} (\boldsymbol{B}_{l})_{\mathcal{M}_{g,f} }$ for all $l \in [L]$ and $f \in [F_g]$, the master can recover the set of blocks $\{\boldsymbol{A}_{g}\boldsymbol{B}_{l}:l\in[L]\}$. 
By executing the processes above for all $g \in[G]$, the master can recover all sets of blocks and outputs $\boldsymbol{A}\boldsymbol{B}$.
Notably, Lagrange codes ensure that the USCTEC scheme tolerates up to $S$ stragglers, since $L+S$ machines in $\mathcal{P}_{g,f}$ are assigned to compute $L+S$ distinct evaluations of the polynomial $H_{g,f}(z)$, while successful decoding requires any $L$ machines. 

It can be seen that in each time step both storage selection and computation assignment, which are determined by $\boldsymbol{\gamma}$ and $\boldsymbol{\mathcal{M}}$, need to be designed. 
In each time step, the system adjust to a corresponding USCTEC scheme, denoted by $(\boldsymbol{\gamma}, \boldsymbol{\mathcal{M}})$.

\subsection{USCTEC with Straggler Tolerance Problem Formulation}
\label{sec-formulation}
For a USCTEC system with a random computation speed $\boldsymbol{\mathsf{s}}$, the goal is to minimize the expected computation time (see Definitions~\ref{def-time} and \ref{def-avertime}). To formulate the problem, we introduce 
the following four definitions.
\begin{definition}
\label{def-matrix}
    {\bf(Load Division Matrix)} For a USCTEC scheme $(\boldsymbol{\gamma}, \boldsymbol{\mathcal{M}})$, the load division matrix is denoted as $\boldsymbol{\mu}\in {\mathbb{R}}^ {G\times N} $. Each entry $\mu[g,n]$ represents the normalized number of columns multiplied by machine $n$ for row block $\boldsymbol{A}_{g}$, i.e., 
\begin{equation}
\label{eq-LDM}
    \mu[g,n]  =
  \begin{cases}
    \frac{\left|\mathcal{D}_{n,g}\right|}{r / L}  & \text{if } n \in \mathcal{U}_{g}, \\
    0  & \text{otherwise, } 
  \end{cases} 
\end{equation}
where $0 \leq \mu[g,n] \leq 1$ for all $g \in [G]$ and $n \in [N]$. 
\end{definition}

Using $\boldsymbol{\mu}$, we can represent $\mathcal{U}_g =  \{n \in [N] : \mu[g,n] >0 \}$ for $g \in [G]$. Hence, from \eqref{eq-storage-selection}, the storage selection $\boldsymbol{\mathcal{I}} = \{\mathcal{I}_n : n\in \mathcal{N}_t \}$ can be represented by the pair $(\boldsymbol{\gamma}, \boldsymbol{\mu})$, where
 \begin{equation}
     \label{eq-store-selection} 
   \mathcal{I}_n  = \{\boldsymbol{A}_g : \mu[g,n] >0,  g \in [G]\}. 
 \end{equation}
\begin{definition}
\label{def-load}
    {\bf(Computation Load)} For a USCTEC scheme $(\boldsymbol{\gamma}, \boldsymbol{\mathcal{M}})$ with a load division matrix $\boldsymbol{\mu}$, the computation load vector is defined as  $\boldsymbol{\theta}$ $=$  $(\theta[1]$, $\cdots$, $\theta[N])$, where $\theta[n] = \sum_{g\in[G]}\gamma[g] \cdot \mu[g,n]$ for $n\in [N]$, i.e., $\boldsymbol{\theta} =\boldsymbol{\gamma} \cdot \boldsymbol{\mu}$. 
\end{definition}
The computation load vector represents the normalized number of row-column multiplications computed by each machine. 

\begin{definition}
    \label{def-time}
    {\bf (Computation Time)}  Given a time step with a computation speed realization $\boldsymbol{s} \in \Omega_{\textbf{s}}$ and a USCTEC scheme $(\boldsymbol{\gamma}, \boldsymbol{\mathcal{M}})$, the computation time is defined as $c(\boldsymbol{\gamma}, \boldsymbol{\mathcal{M}})
           \triangleq  \max_{n\in \mathcal{N}_t} \frac{\theta[n]}{s[n]} = \max_{n\in \mathcal{N}_t} \frac{\sum_{g\in[G]} \gamma[g]\cdot\mu[g,n]}{s[n]}$.
\end{definition}

\begin{definition}
    \label{def-avertime}
     {\bf (Expected Computation Time)} Given a USCTEC system with a speed distribution $\textbf{s}$ and a storage placement $\boldsymbol{\mathcal{Z}}$ that supports a set of USCTEC schemes $\mathcal{T}_{\Omega_{\textbf{s}}} = \{(\boldsymbol{\gamma}, \boldsymbol{\mathcal{M}}) \}$, 
     the expected computation time is defined as $C(\boldsymbol{\mathcal{Z}} , \mathcal{T}_{\Omega_{\textbf{s}}} ) = \mathbb{E}_{\boldsymbol{\mathsf{s}}}\left[ c(\boldsymbol{\gamma}, \boldsymbol{\mathcal{M}})\right]$.
\end{definition}

Our goal is to minimize the expected computation time in Definition \ref{def-avertime} by jointly designing a set of schemes $\mathcal{T}_{\Omega_{\textbf{s}}}$ and the storage placement  $\boldsymbol{\mathcal{Z}}$. We can formulate the following combinatorial optimization problem,
\begin{subequations} 
\label{eq-opt}
\begin{align}
& \underset{\boldsymbol{\mathcal{Z}},\mathcal{T}_{\Omega_{\textbf{s}}}} {\arg\min} \ C(\boldsymbol{\mathcal{Z}},\mathcal{T}_{\Omega_{\textbf{s}}} ) \label{obj}\\
&\text{s.t. } 
0 \leq \frac{|\mathcal{Z}_{n}|}{q} \leq e[n]\leq 1, \ \forall n \in [N],   
\label{eq-store-constraint}\\
&\ \ \ \ \ \forall (\boldsymbol{\gamma}, \boldsymbol{\mathcal{M}}) \in \mathcal{T}_{\Omega_{\textbf{s}}}:  \notag \\
&\ \ \ \ \ \sum_{g \in[G]} \gamma[g] = 1, \ 0 \leq \gamma[g] \leq 1, \ \forall g \in[G],  \label{eq-store-partition}\\
&\ \ \ \ \ \bigcup_{f\in[F_g]}\mathcal{M}_{g,f} = \left[\frac{r}{L}\right],  \forall g \in[G], \label{eq-index-partition}\\
&\ \ \ \ \ \mathcal{P}_{g,f} \subseteq \mathcal{U}_g, \  \forall f \in [F_g], \ g \in [G], \label{eq-select-machine-set}\\
&\ \ \ \ \ |\mathcal{P}_{g,f}| = L+S, \ \forall f \in [F_g], \ g \in [G], \label{eq-select-machine-number}
\end{align}
\end{subequations}
where \eqref{eq-store-constraint} represents storage constraints.
Each USCTEC scheme $(\boldsymbol{\gamma}, \boldsymbol{\mathcal{M}})$ corresponding to a speed realization satisfies constraints \eqref{eq-store-partition}-\eqref{eq-select-machine-number}.
\eqref{eq-store-partition} ensures that each row in matrix $\boldsymbol{A}$ is computed by available machines.
\eqref{eq-index-partition} ensures that each column in $\boldsymbol{B}_{l}$, $l\in [L]$, is assigned to be computed by available machines.
\eqref{eq-select-machine-set} ensures that the assigned machines have stored $\boldsymbol{A}_g$.
\eqref{eq-select-machine-number} ensures that each column is assigned to $L+S$ available machines, providing the straggler tolerance of $S$.

The optimization problem presented in \eqref{eq-opt} is inherently combinatorial, making it challenging to find the optimal solutions. 
In the following sections, we will propose sub-optimal solutions in two steps.
1) We will relax the storage constraint \eqref{eq-store-constraint} by setting $e[n]=1$ for all $n \in  [N]$, and find optimal solutions for a given speed realization. 
2) We will develop a heuristic algorithm for general speed distributions, considering the storage constraint \eqref{eq-store-constraint}.
This algorithm will be based on the approach developed in Step 1). 

\section{Optimal USCTEC Schemes without Storage Constraints for A Given Speed Realization}
\label{sec: no storage constraints}
\subsection{Problem Analysis and An Illustrative Example}
\label{sec-single}
With the relaxed storage constraint $\boldsymbol{e} = \boldsymbol{1}$, where $\boldsymbol{1}$ is an all-$1$ vector, and given a speed realization $\boldsymbol{s}$,
we let machines utilize their entire storage, i.e., $\mathcal{I}_n = \mathcal{Z}_n$ for $n \in \mathcal{N}_t$.
Problem \eqref{eq-opt} is reformulated as the following optimization
problem,
\begin{subequations} 
\label{eq-opt-single}
\begin{align}
& \underset{\boldsymbol{\gamma}, \boldsymbol{\mathcal{M}}} {\arg\min} \ c(\boldsymbol{\gamma}, \boldsymbol{\mathcal{M}}) \label{obj}\\
&\text{s.t. } 
\sum_{g \in[G]} \gamma[g] = 1, \ 0 \leq \gamma[g] \leq 1, \ \forall g \in[G],  \label{eq-store-partition-single}\\
&\ \ \ \ \ \bigcup_{f\in[F_g]}\mathcal{M}_{g,f} = \left[\frac{r}{L}\right],  \forall g \in[G], \label{eq-index-partition-single}\\
&\ \ \ \ \ \mathcal{P}_{g,f} \subseteq \mathcal{U}_g, \  \forall f \in [F_g], \ g \in [G], \label{eq-select-machine-set-single}\\
&\ \ \ \ \ |\mathcal{P}_{g,f}| = L+S, \ \forall f \in [F_g], \ g \in [G]. \label{eq-select-machine-number-single}
\end{align}
\end{subequations}

Based on Definition \ref{def-time}, the computation time $c(\boldsymbol{\gamma}, \boldsymbol{\mathcal{M}})$ is fixed when the computation load vector $\boldsymbol{\theta}$ is fixed.
This insight prompts us to decompose problem \eqref{eq-opt-single} into three sub-problems. First, we solve the optimal computation load vector $\boldsymbol{\theta}^*$ that minimizes the computation time.
Next, we show the existence of a storage placement $\boldsymbol{\mathcal{Z}}^*$, induced by a partitioning vector $\boldsymbol{\gamma}^*$ and a load division matrix $\boldsymbol{\mu}^*$ as shown in \eqref{eq-store-selection}, where $\boldsymbol{\gamma}^* \cdot \boldsymbol{\mu}^* = \boldsymbol{\theta}^*$. 
Finally, we prove the existence of a computation assignment $\boldsymbol{\mathcal{M}}^*$ that satisfies $ \boldsymbol{\mu}^*$. Therefore, an optimal USCTEC scheme $(\boldsymbol{\gamma}^*, \boldsymbol{\mathcal{M}}^*)$ is obtained.
\begin{example}
\label{ex-single}
    When $N = 6$, $L=2$, $S= 1$ and $\boldsymbol{s} =$ $(3$, $3$, $4$, $4$, $5$, $5)$, the optimal computation load vector is $\boldsymbol{\theta}^* =$ $(\frac{3}{8}$, $\frac{3}{8}$, $\frac{1}{2}$, $\frac{1}{2}$, $\frac{5}{8}$, $\frac{5}{8})$, which ensures that all machines complete computing at the same time, resulting in a minimum computation time of $c^* = \frac{1}{8}$. 
    Let $\boldsymbol{\gamma}^* =$ $(\frac{3}{8}$, $\frac{1}{4}$, $\frac{1}{8}$, $\frac{1}{8}$, $\frac{1}{8})$ and
    \begin{small}
    \begin{equation}
    \label{eq-ex-mu}
        \boldsymbol{\mu}^* = 
        \begin{bmatrix}
            1 & 0 & 0 & 0 & 1 & 1 \\
            0 & 0 & 1 & 1 & 1 & 0 \\
            0 & 1 & 1 & 0 & 0 & 1 \\
            0 & 1 & 1 & 1 & 0 & 0 \\
            0 & 1 & 0 & 1 & 0 & 1 
        \end{bmatrix},
    \end{equation}
    \end{small}
    \hspace{-1mm}such that $\boldsymbol{\theta}^* = \boldsymbol{\gamma}^* \cdot\boldsymbol{\mu}^*$. 
    Using $\boldsymbol{\gamma}^*$, the matrix $\boldsymbol{A}$ is divided into $G = 5$ row blocks. Using $\boldsymbol{\mu}^*$, the storage placement from \eqref{eq-store-selection} is as follows. $\mathcal{Z}^*_1 = \{\boldsymbol{A}_1\}$, $\mathcal{Z}^*_2 = \{\boldsymbol{A}_3, \boldsymbol{A}_4,\boldsymbol{A}_5\}$, $\mathcal{Z}^*_3 = \{\boldsymbol{A}_2, \boldsymbol{A}_3, \boldsymbol{A}_4\}$, $\mathcal{Z}^*_4 = \{\boldsymbol{A}_2, \boldsymbol{A}_4, \boldsymbol{A}_5\}$, $\mathcal{Z}^*_5 = \{\boldsymbol{A}_1, \boldsymbol{A}_2\}$ and  
    $\mathcal{Z}^*_6 = \{\boldsymbol{A}_1, \boldsymbol{A}_3,\boldsymbol{A}_5\}$. 
    The sets of selected machines are $\mathcal{U}^*_1 = \{1, 5, 6\}$, $\mathcal{U}^*_2 = \{3, 4, 5\}$, $\mathcal{U}^*_3 = \{2, 3,6\}$, $\mathcal{U}^*_4 = \{2, 3, 4\}$ and $\mathcal{U}^*_5 = \{2, 4, 6\}$.
    Next, we provide a computation assignment $\boldsymbol{\mathcal{M}}^*$. 
    Since $\mu^*[g,n] = 1$ for $n \in \mathcal{U}^*_g$, the indices assigned to each machine $n$ are $\mathcal{D}_{n,g} = [\frac{r}{2}]$ from \eqref{eq-LDM}.
    Since $|\mathcal{U}^*_g| = 3$ and the requirement of $|\mathcal{P}_{g,f}| = 3$ for $f \in [F_g]$, we let $F_g = 1$ for all $g \in [5]$, i.e., $\boldsymbol{\mathcal{M}}^*_g =\{[\frac{r}{2}]\}$ and $\boldsymbol{\mathcal{P}}^*_g = \{\mathcal{U}^*_g\}$. 
    Therefore, we obtain the optimal USCTEC scheme $(\boldsymbol{\gamma}^*, \boldsymbol{\mathcal{M}}^*)$.
\end{example}

We will describe the detailed solution as follows. 


\subsection{Optimal Computation Load Problem}
\label{sec-load}
In this section, we find the optimal computation load.
We introduce the $(l,\boldsymbol{s}, \boldsymbol{\sigma})$-Load Problem, where $\boldsymbol{\sigma}$ is a load constraint vector of length $N$, and $\sigma[n]$ is the maximum load that machine $n \in [N]$ can be assigned.
This problem is used not only for a given speed realization but also for general speed distributions with storage constraints in Section \ref{global}. 
\begin{definition}
    \label{def-p1}
    {\bf ($(l,\boldsymbol{s},\boldsymbol{\sigma})$-Load Problem (LP))} Given $0 \leq l \leq L+S$, a speed realization $\boldsymbol{s}$ and a vector $\boldsymbol{\sigma} =$ $(\sigma[1]$, $\cdots$, $\sigma[N])$, where $l \leq \sum_{n \in \mathcal{N}_t} \sigma[n]$ and $0 \leq \sigma[n] \leq 1$ for all $n \in [N]$, the goal is to find the solution to 
    \begin{subequations}
    \label{eq-opt-load}
    \begin{align}
    & 
    \min_{\boldsymbol{\theta}} \ \max_{n\in \mathcal{N}_t} \frac{\theta[n]}{s[n]} \\
    &\text{s.t.} 
    \sum_{n\in \mathcal{N}_t} \theta[n] = l, \label{eq-sum-load}\\
    &\ \ \ \ \ \ 0 \leq \theta[n] \leq \sigma[n] \leq 1, \ \ \forall n\in \mathcal{N}_t, \label{eq-range-load}\\
    &\ \ \ \ \ \ \theta[n] = 0,  \ \ \forall n \in [N] \setminus \mathcal{N}_t.
    \end{align}
\end{subequations}
\end{definition}
The $(l,\boldsymbol{s}, \boldsymbol{\sigma})$-LP is a convex optimization problem. In fact, its analytical solution can be obtained using Theorem $1$ in \cite{wcj2021hs}.
\begin{theorem}
When $l=L+S$ and $\boldsymbol{\sigma} = \boldsymbol{e} = \boldsymbol{1}$, the optimal computation load vector, induced by the solution to problem \eqref{eq-opt-single}, is the solution to $(L+S, \boldsymbol{s}, \boldsymbol{1})$-LP, without considering an explicit storage placement and computation assignment.
\end{theorem}
\begin{IEEEproof}
Given the optimal solution to problem \eqref{eq-opt-single}, \eqref{eq-range-load} is satisfied, due to $\theta[n]  = \sum_{g\in[G]}\gamma[g] \cdot \mu[g,n] = $ $ \sum_{g\in[G]: \mu[g,n] > 0} \gamma[g] \cdot \mu[g,n] \overset{(a)}{\leq} $ $\sum_{g\in[G]: \mu[g,n] > 0} \gamma[g] \overset{(b) }{=} \frac{|\mathcal{Z}_n|}{q}$ $\leq e[n]$ $=1$, where $(a)$ is due to $\mu[g,n] \leq 1$ from \eqref{eq-LDM}, and $(b)$ is due to \eqref{eq-store-selection}.
To show \eqref{eq-sum-load},
we first claim the following  constraint of the load division matrix,
\begin{equation}
\label{eq-transform}
    \sum_{n \in[N]} \mu[g,n] = L+S 
\end{equation}
for $g \in [G]$. This is due to 
$\sum_{n \in[N]} \mu[g,n]$ $\overset{(a)}{=}$
$ \frac{\sum_{n \in \mathcal{U}_g} |\mathcal{D}_{n,g}|}{r/L} =$ $ \frac{\sum_{n \in \mathcal{U}_g} \sum_{f\in[F_g]: n \in \mathcal{P}_{g,f}}|\mathcal{M}_{g,f}|}{r/L}$ $\overset{(b)}{=}$ $\frac{\sum_{f \in [F_g]} \sum_{n \in \mathcal{P}_{g,f}}|\mathcal{M}_{g,f}|}{r/L}$ $\overset{(c)}{=}$ $\frac{\sum_{f \in [F_g]} (L+S)\cdot |\mathcal{M}_{g,f}|}{r/L}$ $\overset{(d)}{=}$ $L+S$, where $(a)$ is due to \eqref{eq-LDM},  $(b)$ is due to \eqref{eq-select-machine-set-single}, $(c)$ is due to \eqref{eq-select-machine-number-single} and $(d)$ is due to \eqref{eq-index-partition-single}.
Hence, $\sum_{n \in \mathcal{N}_t} \theta [n]$ $=$ $\sum_{n \in [N]}$ $ \sum_{g \in [G]}$ $\gamma[g]\mu[g,n]=$ $\sum_{g \in [G]}$ $ \left(\gamma[g] \cdot \sum_{n\in [N]} \mu[g,n] \right)$ $= \sum_{g \in [G]} \gamma[g] \cdot (L+S) = L+S$. 
\end{IEEEproof}


\subsection{Storage Placement Problem}
\label{sec-division}
To obtain a partitioning vector $\boldsymbol{\gamma}$ and a load division matrix $\boldsymbol{\mu}$, given a load vector $\boldsymbol{\theta}$, we introduce the $(\boldsymbol{\theta}, \rho)$-Division Problem, where $\rho$ is the sum of $\boldsymbol{\gamma}$ and represents a fraction of the data matrix $\boldsymbol{A}$ to be partitioned. In problem \eqref{eq-opt-single}, we consider $\rho = 1$, while $\rho \neq 1$ will be used in  Section \ref{global}.
\begin{definition}
    \label{def-p2} 
    {\bf (($\boldsymbol{\theta}, \rho)$-Division Problem (DP))} Given a computation load vector $\boldsymbol{\theta} \in \mathbb{R}^{N}$ and $0 \leq \rho \leq 1$, where $\sum_{n \in [N]}\theta[n]$ $= (L+S)\rho$ and $0 \leq \theta[n] \leq \rho$, the goal is to find a vector $\boldsymbol{\gamma} \in \mathbb{R}^{G}$ and a matrix $\boldsymbol{\mu} \in {\mathbb{R}}^{G\times N}$ such that 
\begin{subequations} 
\label{eq-opt-dp}
\begin{align}
&{\boldsymbol{\theta}} = \boldsymbol{\gamma}\cdot \boldsymbol{\mu}, \label{eq-dp-product} \\
&\sum_{g \in[G]} \gamma[g] = \rho, \ 0 \leq \gamma[g] \leq 1,  \ \forall g \in[G],  \label{eq-dp-fraction} \\
&\sum_{n\in[N]} \mu[g,n] = L+S, \ \forall g \in[G],  \label{eq-dp-sum}\\
& 0 \leq \mu[g,n] \leq  1, \ \ \forall n\in \mathcal{N}_t, g\in[G], \label{eq-dp-range}\\
& \mu[g,n] = 0, \ \forall n \in [N] \setminus \mathcal{N}_t. \label{eq-dp-0}
\end{align}
\end{subequations}
\end{definition}

\begin{theorem}
\label{th-gm}
   The solution to $(\boldsymbol{\theta}^*, 1)$-DP consists of the partitioning vector and load division matrix induced by the optimal solution to problem \eqref{eq-opt-single},
   without considering an explicit computation assignment $\boldsymbol{\mathcal{M}}$, where $\boldsymbol{\theta}^*$ is the optimal computation load obtained from $(L+S,\boldsymbol{s},\boldsymbol{1})$-LP.
\end{theorem}

\begin{IEEEproof}
    For any solution to $(\boldsymbol{\theta}^*, 1)$-DP, i.e., $\boldsymbol{\gamma}^*$ and $\boldsymbol{\mu}^*$, we let $\boldsymbol{\gamma}^*$ be the partitioning vector in problem \eqref{eq-opt-single}, as \eqref{eq-store-partition-single} is satisfied from \eqref{eq-dp-fraction}. Let $\boldsymbol{\mu}^*$ be the load division matrix induced by the solution to problem \eqref{eq-opt-single}, as \eqref{eq-transform} is satisfied from \eqref{eq-dp-sum}. From \eqref{eq-dp-product}, 
    any computation assignment satisfying $\boldsymbol{\mu}^*$ achieves the optimal computation time.
\end{IEEEproof}

To derive a solution to $(\boldsymbol{\theta}, \rho )$-DP, 
we specify \eqref{eq-dp-range} as $\mu[g,n] = 1$ or $0$, such that the desired binary matrix $\boldsymbol{\mu}$ contains $L+S$ ``$1$''s in each row. 
We denote the specified problem as Binary-$(\boldsymbol{\theta}, \rho )$-DP, which is a Filling Problem introduced in \cite{wcj2019fp}. 
Lemma \ref{le-existence} provides the necessary and sufficient conditions for a solution exist in Binary-$(\boldsymbol{\theta}, \rho)$-DP.
\begin{lemma}
    \label{le-existence}
    (\!\!\cite{wcj2019fp}) 
    The solution to Binary-$(\boldsymbol{\theta}, \rho)$-DP exists if and only if $\theta[n] \leq \frac{\sum_{i \in [N]} \theta[i]}{L+S}$
    for all $n \in [N]$.
\end{lemma}
From Lemma \ref{le-existence}, there always exist solutions to  Binary-$(\boldsymbol{\theta}, \rho )$-DP, due to $\theta[n] \leq \rho = \frac{\sum_{i \in [N]} \theta[i]}{L+S} $ for $n \in [N]$.
\begin{solution-to-dp}
\label{so-1}
For Binary-$(\boldsymbol{\theta}, \rho )$-DP, we present $(\boldsymbol{\theta}, \rho)$-Division Algorithm, by generalizing the algorithm in \cite{wcj2019fp} using a scalar $0 \leq \rho \leq 1$, which originally considers $\rho = 1$. With the input $\boldsymbol{\theta}$ and $\rho$, we obtain outputs $\boldsymbol{\gamma}$ and $\boldsymbol{\mu}$ as shown in Algorithm~\ref{al-cec}, which are the solution to Binary-$(\boldsymbol{\theta}, \rho )$-DP.
\end{solution-to-dp}

\floatstyle{spaceruled}
\restylefloat{algorithm}
\begin{algorithm}[t]
  \caption{$(\boldsymbol{\theta}, \rho)$-Division Algorithm}
  \label{al-cec}
  \begin{algorithmic}[1]
  \item [{\bf Input}: $\boldsymbol{\theta}$, $\rho$]
   \hspace*{4cm} 
    \STATE $g \leftarrow 0$
    \WHILE{$\boldsymbol{\theta}$ contains a non-zero element}
        \STATE $g \leftarrow g+1$
        \STATE $L' \leftarrow \sum_{i=1}^{N} \theta[i]$
        \STATE $N' \leftarrow $ number of non-zero elements in $\boldsymbol{m}$
        \STATE $\boldsymbol{o} \leftarrow $ indices that sort the non-zero elements of $\boldsymbol{\theta}$ in ascending order
        \STATE $\mathcal{U}_{g} \leftarrow \{o[1], o[N'-(L+S)+2], \cdots, o[N'] \}$
        \STATE $\boldsymbol{b}_g \leftarrow$ a $\{0, 1\}$-vector where $b_g[i] = 1$ if $i \in \mathcal{U}_g$ 
        \IF{$N' \geq L+S+1$}
            \STATE $\gamma_{g} \!\leftarrow  \!\!\frac{1}{\rho} \min \left(
            \!\!\frac{L'}{L+S}\! - \theta\left[o[N'-\!(L+S)\!+\!1]\right], \theta[o[1]]\right)$
        \ELSE
            \STATE $\gamma_{g} \leftarrow \theta[o[1]]\cdot \frac{1}{\rho}$
        \ENDIF
        \FOR{$n \in \mathcal{U}_{g}$ }
            \STATE $\theta[n] \leftarrow \theta[n] - \gamma_{g}\rho$
        \ENDFOR
    \ENDWHILE
    \STATE $G \leftarrow g$
    \STATE $\boldsymbol{\gamma} \leftarrow$ a vector of length $N$, where $\gamma[g] = \gamma_{g}\cdot \rho$ for $g \in [G]$
    \STATE $\boldsymbol{\mu} \leftarrow$ a matrix of size $G\times N$, where $\boldsymbol{\mu}[g] = \boldsymbol{b}_g$ for $g \in [G]$  
  \item [{\bf Output }: $\boldsymbol{\gamma}$, $\boldsymbol{\mu}$ ]
  \end{algorithmic}
\end{algorithm}




\subsection{Computation Assignment Problem}
\label{sec-assignment}
Given any load division matrix $\boldsymbol{\mu}$ of size $G\times N$, 
designing a computation assignment $(\boldsymbol{\mathcal{M}}_g,\boldsymbol{\mathcal{P}}_g)$ for $g \in [G]$ is equivalent to solving a Binary-$(\boldsymbol{\mu}[g], 1)$-DP, by two steps as follows. 
For clarity, we denote the desired vector and matrix in Binary-$(\boldsymbol{\mu}[g], 1)$-DP as $\boldsymbol{\gamma}'$ and $\boldsymbol{\mu}'$, respectively. 
First, we let $F_g = |\boldsymbol{\gamma}'|$ and partition the indices $[\frac{r}{L}]$ into $F_g$ disjoint sets $\mathcal{M}_{g,1}$, $\cdots$, $\mathcal{M}_{g,F_g}$ of size $\frac{\gamma'_1 \cdot r} {L}$, $\cdots$, $\frac{\gamma'_{F_g} \cdot r}{L}$ respectively. 
Second, we let $\mathcal{M}_{g,f} = \{n : \mu'[f,n]  = 1\}$ for $f \in [F_g]$. 
From \eqref{eq-dp-product}, the obtained $(\boldsymbol{\mathcal{M}}_g,\boldsymbol{\mathcal{P}}_g)$ satisfies vector $\boldsymbol{\mu}[g]$, i.e., $\boldsymbol{\gamma}' \cdot \boldsymbol{\mu}' = \boldsymbol{\mu}[g]$.
Moreover, there always exist solutions to Binary-$(\boldsymbol{\mu}[g], 1)$-DP, as $\mu[g,n] \leq 1 = \frac{\sum_{i \in [N]}\mu[g,i]}{L+S}$ for all $n \in [N]$, satisfying the condition in Lemma \ref{le-existence}. 
Therefore, we obtain $(\boldsymbol{\mathcal{M}}_g,\boldsymbol{\mathcal{P}}_g)$ for $g \in [G]$ by solving the Binary-$(\boldsymbol{\mu}[g], 1)$-DP using Solution \ref{so-1}, and using two steps as discussed. 

\section{General Solutions for USCTEC with Storage Constraints} 
\label{global}
Algorithm~\ref{al-storage} provides a general solution for USCTEC systems with storage constraints, by generating a storage placement $\boldsymbol{\mathcal{Z}}$ and storage selections for a general speed distribution. A detailed illustration is provided in Example \ref{ex-general}.
The idea is to unionize the storage selections for all speed realizations. 
However, if the combined storage  exceeds the storage constraint of any machine, it results in a \emph{storage overflow}. 
In such cases, the machines with storage overflow will fill their storage capacity, and the storage placement will be adjusted for the remaining machines in a similar fashion.

\floatstyle{spaceruled}
\restylefloat{algorithm}
\begin{algorithm}
  \caption{Storage Placement and Storage Selections}
  \label{al-storage}
  \begin{algorithmic}[1]
  \item[{\bf Input}: $\Omega_{\textbf{s}}$, $N$, $L$, $S$ ]
  \hspace*{4cm} 
  \STATE $\hat{\rho} \leftarrow 0$ 
 \STATE $\hat{\boldsymbol{\gamma}}_{\boldsymbol{s}} \leftarrow \boldsymbol{0}$ of length $1$, $\hat{\boldsymbol{\mu}}_{\boldsymbol{s}} \leftarrow \boldsymbol{0}$ of size $1\times N$, $\boldsymbol{\sigma}_{\boldsymbol{s} } \leftarrow \boldsymbol{1}$ of length $N$, and $l_{\boldsymbol{s}} \leftarrow L+S$ for all $\boldsymbol{s} \in \Omega_{\textbf{s} }$
  \WHILE{$\sum_{\boldsymbol{s} \in \Omega_{\textbf{s}}} l_{\boldsymbol{s}} >0$}
    \STATE $\mathcal{Z}_{n} \leftarrow  \emptyset $ for $n \in[N]$  
    \FOR {$\boldsymbol{s} \in \Omega_{\textbf{s}}$} 
        \STATE $\boldsymbol{\bar{\theta}}_{\boldsymbol{s}} \leftarrow $ solution to the $\left( l_{\boldsymbol{s}}, \boldsymbol{s}, \boldsymbol{\sigma}_{\boldsymbol{s}} \right)$-LP
        \STATE $(\boldsymbol{\bar{\gamma}}_{\boldsymbol{s}}, \boldsymbol{\bar{\mu}}_{\boldsymbol{s}}) \leftarrow$ solution to the $(\boldsymbol{\bar{\theta}}_{\boldsymbol{s}}, 1 -\hat{\rho})$-DP
        \STATE $\left(\boldsymbol{\gamma}_{\boldsymbol{s}}, \boldsymbol{\mu}_{\boldsymbol{s}} \right) \leftarrow \left([\hat{\boldsymbol{\gamma}}_{\boldsymbol{s}}, \boldsymbol{\bar{\gamma}}_{\boldsymbol{s}} ], 
        \begin{bmatrix} \hat{\boldsymbol{\mu}}_{\boldsymbol{s}} \\ \boldsymbol{\bar{\mu}}_{\boldsymbol{s}} \end{bmatrix} \right)$
        \STATE $\boldsymbol{\mathcal{I}}_{\boldsymbol{s}} \leftarrow $ the storage selection based on $(\boldsymbol{\gamma}_{\boldsymbol{s}}, \boldsymbol{\mu}_{\boldsymbol{s}})$ 
        \STATE $\mathcal{Z}_{n} \leftarrow \mathcal{Z}_{n} \bigcup \mathcal{I}_{\boldsymbol{s},n}$ for $n \in [N]$
    \ENDFOR
    \IF{there exists storage overflow on $\mathcal{Z}_n$, $n \in [N]$}
        \STATE $\hat{\rho} \leftarrow $ the location of the first row that overflows
        \FOR{$\boldsymbol{s} \in \Omega_{\textbf{s}}$} 
            \STATE $\left(\hat{\boldsymbol{\gamma}}_{\boldsymbol{s}}, \hat{\boldsymbol{\mu}}_{\boldsymbol{s}} \right) \leftarrow \left({\boldsymbol{\gamma}_{\boldsymbol{s}}}_{\prec \hat{\rho}}, {\boldsymbol{\mu}_{\boldsymbol{s}}}_{\prec \hat{\rho}} \right)$
            \STATE $l_{\boldsymbol{s}} \leftarrow (L+S)(1-\hat{\rho})$ 
            \STATE $s[n] \leftarrow 0$ for $n$ that has storage overflow on $\hat{\rho}$
            \STATE $\boldsymbol{\sigma}_{\boldsymbol{s}} \leftarrow (1-\hat{\rho},\cdots, 1-\hat{\rho}) $  of length $N$
        \ENDFOR
    \ELSE
        \STATE $\boldsymbol{\mathcal{I}}_{\boldsymbol{s}} \leftarrow$ the storage selection for $\boldsymbol{s} \in  \Omega_{\textbf{s}}$ 
    \ENDIF
  \ENDWHILE
  \item[{\bf Output}: $\boldsymbol{\mathcal{Z}}$, $\{\boldsymbol{\mathcal{I}}_{\boldsymbol{s}}  : \boldsymbol{s} \in  \Omega_{\textbf{s}}\}$]
  \end{algorithmic}
\end{algorithm}

\begin{example}
\label{ex-general}
Consider a system with $N=6$, $L=2$, $S=1$, $\boldsymbol{e} =$ $(0.6$, $0.6$, $0.8$, $0.8$, $1$, $1)$, two speed realizations $\boldsymbol{s}_1 =$ $(3$, $3$, $4$, $4$, $5$, $5)$ and $\boldsymbol{s}_2 =$ $(3$, $1$, $2$, $2$, $3$, $5)$ with equal probabilities.
The locations of rows in data matrix $\boldsymbol{A}$ are represented by real numbers in the range $[0,1]$. Specifically, the $aq$-th row is located at $a$. We simplify all notations $(\cdot)_{\boldsymbol{s}_i}$ in Algorithm \ref{al-storage} as $(\cdot)_{i}$ for $i \in [2]$. For example, we simplify $\boldsymbol{\gamma}_{\boldsymbol{s}_i}$ as $\boldsymbol{\gamma}_{i}$.

{\bf Optimal USCTEC Schemes without Storage Constraints (Lines $6$-$9$) :} For $\boldsymbol{s}_i$, $i \in [2]$, we obtain the partitioning vector $\boldsymbol{\bar{\gamma}}_i$, load division matrix $\boldsymbol{\bar{\mu}}_i$ by solving problems shown in lines $6$ and $7$. In line $8$, we have $\boldsymbol{\gamma}_{i} = [\boldsymbol{0}, \bar{\boldsymbol{\gamma}}_i]$ and 
$\boldsymbol{\mu}_i 
= \begin{bmatrix}
    \boldsymbol{0} \\
    \bar{\boldsymbol{\mu}_i}
\end{bmatrix}$. We simplify them as $\boldsymbol{\gamma}_{i} = \bar{\boldsymbol{\gamma}}_i$ and $\boldsymbol{\mu}_i =  \bar{\boldsymbol{\mu}_i}$.
Specifically, 
$\boldsymbol{\gamma}_1 =$ $(\frac{3}{8}$, $\frac{1}{4}$, $\frac{1}{8}$, $\frac{1}{8}$, $\frac{1}{8})$, $\boldsymbol{\gamma}_2  =$ $(\frac{3}{16}$, $\frac{3}{8}$, $\frac{1}{16}$, $\frac{1}{16}$, $\frac{1}{16}$, $\frac{1}{4})$, $\boldsymbol{\mu}_1$ is shown in \eqref{eq-ex-mu}. 
In line $9$, we obtain the storage selection $\boldsymbol{\mathcal{I}}_{i} =$ $\{\mathcal{I}_{i,n} : n\in [6]\}$, where 
$\mathcal{I}_{i,n}$ ($\mathcal{I}_{\boldsymbol{s}_i,n}$), is the storage selection of machine $n$.

{\bf Storage Overflow (Lines $10$-$13$):} If we use $\mathcal{I}_{1,n}\bigcup\mathcal{I}_{2,n}$ 
as the storage placement for machine $n \in [6]$, a storage overflow occurs with machine $1$ at the row located at $\frac{3}{5}$. In this case, we first define the storage placement and storage selections for rows in $[0, \frac{3}{5})$, and then reassign rows in $[\frac{3}{5},1]$.

{\bf Assign Rows in $[0,\frac{3}{5})$ (Lines $14$-$19$):}
Each machine $n$ $\in [6]$ stores rows in $\mathcal{I}_{1,n} \bigcup \mathcal{I}_{2,n}$ subsequently, until they reach the row located at $\frac{3}{5}$. 
Correspondingly, we modify partitioning vectors and load division matrices, as shown in line $15$.
For each $\boldsymbol{s}_i$, $i \in [2]$, we truncate $\boldsymbol{\gamma}_i$ to obtain a shorter vector with a sum of $ \hat{\rho} = \frac{3}{5}$, denoted by $\boldsymbol{\gamma}_{i_{\prec \hat{\rho}}}$. We then obtain
$\boldsymbol{\gamma}_{1_{\prec \hat{\rho}}} = (\frac{3}{8}, \frac{9}{40})$ and $\boldsymbol{\gamma}_{2_{\prec \hat{\rho}}} = (\frac{3}{16}, \frac{3}{8}, \frac{3}{80})$ with length of $2$ and $3$, respectively.
We truncate $\boldsymbol{\mu}_1$ to $\boldsymbol{\mu}_{1_{\prec \hat{\rho}}}$ with  $2$ rows, and truncate $\boldsymbol{\mu}_2$ to $\boldsymbol{\mu}_{2_{\prec \hat{\rho}}}$ with $3$ rows, where
\begin{small}
\begin{equation*}
    \boldsymbol{\mu}_{1_{\prec \hat{\rho}}} = 
\begin{bmatrix}
1 & 0 & 0 & 0 & 1 & 1\\
0 & 0 & 1 & 1 & 1 & 0 
\end{bmatrix}, 
\boldsymbol{\mu}_{2_{\prec \hat{\rho}}} =
\begin{bmatrix}
0 & 1 & 0 & 0 & 1 & 1 \\
1 & 0 & 1 & 0 & 0 & 1 \\
1 & 0 & 0 & 0 & 1 & 1
\end{bmatrix}.
\end{equation*}
 \end{small}
For $\boldsymbol{s}_i$, $i \in [2]$, the remaining load is $(L+S)(1-\hat{\rho}) = \frac{6}{5}$.
We update $\boldsymbol{s}_i$ to $\boldsymbol{s}'_i$, where $s'_i[n] = 0$ if $n = 1$, otherwise $s'_i[n] = s_i[n]$, and update $\boldsymbol{\sigma}_i$ to $\boldsymbol{\sigma}'_i  = (1- \hat{\rho}, \cdots, 1-\hat{\rho})$.

{\bf Reassign the Rows in $[\frac{3}{5},1]$ (Lines $5$-$11$):}
For each $\boldsymbol{s}_i$, $i \in [2]$, we solve the $(\frac{6}{5}, \boldsymbol{s}'_i, \boldsymbol{\sigma}'_i)$-LP to obtain load vector $\boldsymbol{\bar{\theta}}'_i$, where $\boldsymbol{\bar{\theta}}'_1 =$ $(0$, $\frac{6}{35}$, $\frac{8}{35}$, $\frac{8}{35}$, $\frac{2}{7}$, $\frac{2}{7})$ and $\boldsymbol{\bar{\theta}}'_2 =$ $(0$, $\frac{1}{10}$, $\frac{1}{5}$, $\frac{1}{5}$, $\frac{3}{10}$, $\frac{2}{5})$. 
We solve the Binary-$(\boldsymbol{\bar{\theta}}'_i, 1- \hat{\rho})$-DP to obtain a vector $\boldsymbol{\bar{\gamma}}'_i$ and a matrix $\boldsymbol{\bar{\mu}}'_i$. The setting of load constrains $\boldsymbol{\sigma}'_i$ is to ensure that the obtained $\boldsymbol{\bar{\theta}}'_i$ satisfies the condition in Lemma \ref{le-existence}, such that Binary-$(\boldsymbol{\bar{\theta}}'_i, 1- \hat{\rho})$-DP has solutions. 
Specifically, $\boldsymbol{\bar{\gamma}}'_1  =$ $(\frac{43}{250}$, $\frac{57}{500}$, $\frac{57}{500})$, $\boldsymbol{\bar{\gamma}}'_2  =$ $(\frac{1}{10}$, $\frac{1}{10}$, $\frac{1}{10}$, $\frac{1}{10})$, 
\begin{small}
\begin{equation*}
    \bar{\boldsymbol{\mu}}'_1 =
\begin{bmatrix}
    0 & 1 & 0 & 0 & 1 & 1 \\
    0 & 0 & 1 & 1 & 1 & 0 \\
    0 & 0 & 1 & 1 & 0 & 1
\end{bmatrix} \hbox{ and }
\bar{\boldsymbol{\mu}}'_2 =
\begin{bmatrix}
    0 & 1 & 0 & 0 & 1 & 1 \\
    0 & 0 & 0 & 1 & 1 & 1 \\
    0 & 0 & 1 & 0 & 1 & 1 \\
    0 & 0 & 1 & 1 & 0 & 1 
\end{bmatrix}.
\end{equation*}  
\end{small}
As shown in line $8$,  we consider the combined partitioning vectors and load division matrices, i.e.,  
$\boldsymbol{\gamma}'_1 =$  $[\boldsymbol{\gamma}_{1_{\prec \hat{\rho}}}$,  $\bar{\boldsymbol{\gamma}}'_1 ]$= $(\frac{3}{8}$, $\frac{9}{40}$, $\frac{43}{250}$, $\frac{57}{500}$, $\frac{57}{500})$, 
$\boldsymbol{\gamma}'_2 =$  $[\boldsymbol{\gamma}_{2_{\prec \hat{\rho}}}$, $\boldsymbol{\bar{\gamma}}'_2 ]$ = $(\frac{3}{16}$, $\frac{3}{8}$, $\frac{3}{80}$, $\frac{1}{10}$, $\frac{1}{10}$, $\frac{1}{10}$, $\frac{1}{10})$, 
$\boldsymbol{\mu}'_1 =
     \begin{bmatrix}
        \boldsymbol{\mu}_{1_{\prec \hat{\rho}}} \\
         \boldsymbol{\bar{\mu}}'_1
    \end{bmatrix}$ and 
$\boldsymbol{\mu}'_2 = 
\begin{bmatrix}
    \boldsymbol{\mu}_{2_{\prec \hat{\rho}}}  \\
    \boldsymbol{\bar{\mu}}'_2
\end{bmatrix}$.   
From \eqref{eq-store-selection}, we obtain the storage selection $\boldsymbol{\mathcal{I}}_i$ for $\boldsymbol{s}_i$, using $(\boldsymbol{\gamma}'_i, \boldsymbol{\mu}'_i)$, where $i \in [2]$.

{\bf Storage Placement and Storage Selections (Line $21$):}
It can be seen that there is no storage overflow, by letting the storage placement for machine $n$ be $\mathcal{Z}_n = \mathcal{I}_{1,n}\bigcup\mathcal{I}_{2,n}$. Therefore, the storage placement $\boldsymbol{\mathcal{Z}}= \{\mathcal{Z}_n: n\in [6]\}$, storage selections $\boldsymbol{\mathcal{I}}_1$ and $\boldsymbol{\mathcal{I}}_2$ for the system are obtained, which are visualized in Fig. \ref{fig-union-with}.

\begin{figure}
\centering
\includegraphics[width=3.2in]{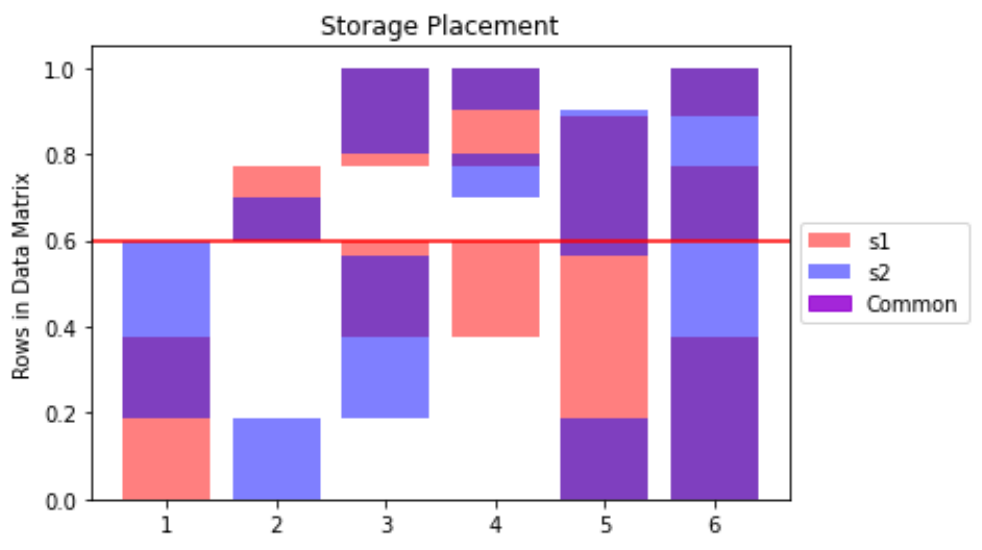}
\caption{Storage Placement and Storage Selections in Example \ref{ex-general}: The $x$-axis represents machine labels. The $y$-axis represents the location of rows in $\boldsymbol{A}$. The union of red and purple bars represents the storage selection for $\boldsymbol{s}_1$. The union of blue and purple bars represents the storage selection for $\boldsymbol{s}_2$. The purple bars represent the common storage selection for both $\boldsymbol{s}_1$ and $\boldsymbol{s}_2$. The red line at $y = \frac{3}{5}$ indicates a storage overflow occurred on machine $1$. }
\label{fig-union-with}
\end{figure}
\end{example}

\section{Discussions}
\label{sec-com-cyc}
We compare Algorithm \ref{al-storage} with USCTEC systems based on cyclic storage strategy presented in \cite{usutec2022}.
We use the following example to compare the storage size and expected computation time obtained by two USCTEC systems.

Consider a system with  $N=12$, $L=2$, $S=1$, and two speed realizations $\boldsymbol{s}_1$ and $\boldsymbol{s}_2$ with equal probabilities, where $\boldsymbol{s}_1 = (1$, $1$, $2$, $2$, $2$, $3$, $8$, $8$, $8$, $8$, $9$, $9)$ and $\boldsymbol{s}_2 = (8$, $8$, $2$, $3$, $9$, $9$, $2$, $1$, $8$, $5$, $2$, $8)$.
We define the storage constraint as $\boldsymbol{e} = (\frac{Q}{12},\cdots, \frac{Q}{12}) $ of length $12$, where $Q \in \{6, 7, 8, 9, 10, 11, 12\}$.
The USCTEC system based on cyclic storage placement \cite{usutec2022} operates as follows.
First, each machine utilizes the full storage capacity by defining $\boldsymbol{\gamma} = (\frac{1}{12}, \cdots, \frac{1}{12})$ of length $12$, and letting the $n$-th machine store $Q$ blocks $\boldsymbol{A}_{n \% N}$, $\cdots$, $\boldsymbol{A}_{(n+Q-1)\%N}$, where we define $a \% N \triangleq a - \lfloor  \frac{a-1}{N}\rfloor N$.
Second, it can be shown that, given the storage placement, the system achieves the minimum computation time. Specifically,
For $g \in [12]$,  $\mathcal{U}_g$ is the set of all machines that store block $\boldsymbol{A}_g$, and $\boldsymbol{\mu}[g]$ is the solution to $(L+S, \boldsymbol{s}_{\mathcal{U}_g}, \boldsymbol{1})$-LP, where $\boldsymbol{s}_{\mathcal{U}_g}$ is a vector containing the computation speeds of machines in $\mathcal{U}_g$. 
By varying storage constraints, we have comparisons as shown in Table \ref{table-com-cyc}.
    \begin{table}[H]
    \caption{Comparisons to Cyclic Storage Placement \label{table-com-cyc}}
    \centering
    \begin{tabular}{|c|c|c|c|c|c|c|}
    \hline
                         &   \multicolumn{2}{c|}{Cyclic Storage Placement}   &  \multicolumn{2}{c|}{Algorithm \ref{al-storage}} \\
    \hline
    $\frac{Q}{N}$ & Storage Size & $C(\boldsymbol{\mathcal{Z}} , \mathcal{T}_{\Omega_{\textbf{s}}} )$  &  Storage Size & $C(\boldsymbol{\mathcal{Z}} , \mathcal{T}_{\Omega_{\textbf{s}}} )$  \\
    \hline
    $\frac{6}{12}$ & $6$ & $0.07235$ & $5.16591$ & $0.09164$ \\
    \hline
    $\frac{7}{12}$ & $7$ & $0.06072$ & $5.23310$ & $0.04812$ \\
    \hline
    $\frac{8}{12}$ & $8$ & $0.05371$ & $5.23480$ & $0.04766$ \\
    \hline
    $\frac{9}{12}$ & $9$ & $0.05101$ & $5.23480$ & $0.04766$ \\
    \hline
    $\frac{10}{12}$ & $10$ & $0.04927$ & $5.23480$ & $0.04766$ \\
    \hline
    $\frac{11}{12}$ & $11$ & $0.04812$ & $5.23480$ & $0.04766$ \\
    \hline
    $\frac{12}{12}$ & $12$ & $0.04766$ & $5.23480$ & $0.04766$  \\
    \hline
    \end{tabular}
\end{table} 
From Table~\ref{table-com-cyc}, it can be seen that 
the proposed algorithm achieves a smaller storage size compared to the baseline algorithm. In addition, except the case when the storage constraint is $\frac{1}{2}$, the achieved expected computation time of the proposed algorithm is always smaller than or equal to the baseline algorithm.
In particular, as the storage constraint increases to $\frac{2}{3}$ and larger, we can show that the systems using Algorithm \ref{al-storage} 
achieve the optimal expected computation time of $0.04766$ and a storage size of $5.23480$.

\bibliographystyle{IEEEtran}
\bibliography{reference}

\end{document}